\documentclass{appolb}
\usepackage{epsfig}
\usepackage{amssymb}

\begin{document}

\title{The significance of using the Newcomb\,--\,Benford law as a test of nuclear half-life calculations}
\author{J\'anos~Farkas, Gy\"orgy~Gy\"urky
\address{Institute of Nuclear Research (ATOMKI), H-4001 Debrecen, POB.\,51, Hungary}
}
\maketitle


\begin{abstract}
Half-life number sequences collected from nuclear data charts are found to obey the Newcomb\,--\,Benford law. Based on this
fact, it has been suggested recently, that this law should be used to test the quality of nuclear decay
models. In this paper we briefly recall how, when and why the Newcomb\,--\,Benford law can be observed in a set of numbers with
a given probability distribution. We investigate the special case of nuclear half-lives, and show that the law provides no
additional clue in understanding decay half-lives. Thus, it can play no significant role in testing
nuclear decay theories.
\end{abstract}

  
\section{Introduction}\label{sec:1}
The history of the Newcomb\,--\,Benford law (NBL) goes back to the nineteenth century, when Newcomb found a rather strange `law' of
nature: the end of the logarithmic tables are much less used than their beginning \cite{Newcomb}. The law had been
reinvented later by Frank Benford, who extended it by investigating an enormous quantity of data from various
 sources \cite{Benford}. To each element of a given data sequence he mapped their first significant digit, and
plotted the histogram of the new sequence. In most of the cases he found the same strange pattern: the probability
$P_d$ of a number having $d$ as its first significant digit follows the law we now call Benford's first digit law:
\begin{equation} \label{eq:BL}
P_d = \lg\big(1+\frac{1}{d}\big) \qquad \qquad (d = 1,\,2,\,\dots,\,9).
\end{equation}

\par We note, that the NBL is more general and covers the other digits, too \cite{Newcomb}. In spite of this,
only the first
digit law can be examined in connection with nuclear half-lives. The main reason for this is that in many cases we
do not know the second digit due to experimental errors. What is more, we only know a few thousand half-life values,
which does not give enough statistics to provide reliable results on the distribution of other digits.

\par The NBL was found to be base invariant, meaning that if the numbers in the original sequence obeying the
law is transformed to a logarithmic base of $k$, and the first significant digits are extracted in this new base, then the law
still holds:
\[ P_d = \log_k\big(1+\frac{1}{d}\big) \qquad \qquad (d = 1,\,\dots,\,k-1). \]
The only obvious limitation is that $k$ cannot be arbitrarily big.

\par More excitingly, if one has the elements of a sequence obeying the NBL multiplied by a given
constant, the new sequence will again follow the law. This means that the NBL is scale invariant.

\par No matter if the primary data sequence comes from the using of logarithmic tables or numbers in wealth
statistics, magazines, geographic data or physics books, there is a very good chance that they will obey the law.
The validity of the NBL for different kind of number sequences is a long standing issue in mathematics and the
natural sciences. In physics, there are many sequences fulfilling the law approximately well, like physical
constants \cite{Burke91}, seismic activity data \cite{Pietronero01} or the strengths of the lines in atomic spectra 
\cite{Pain08}. The half-lives of $\alpha$-radioactive nuclei has also been examined \cite{Buck93}, and
most recently this investigation has been extended to $\beta$ decay and spontaneous fission, including
also ground state and isomeric nuclei \cite{NiRen08, NiRen09}.

\par In this paper, we examine the recent statements and conclusions regarding the connection between the NBL and nuclear decay \cite{Buck93, NiRen08, NiRen09} in the light of the mathematical explanation of the law.


\section{The chart of half-lives} \label{sec:2}

\subsection{The distribution of the first digits} \label{sec:2.1}

\par In Refs.~\cite{Buck93,NiRen08,NiRen09}, the authors examine the relative occurrence of the first
significant decimal
digits of nuclear half-life values. The data are found to satisfy the NBL. In \cite{Buck93},
477 $\alpha$ decaying nuclei are taken into account, while in \cite{NiRen09}, 2059 $\beta$ decay
half-lives are investigated. Ref.~\cite{NiRen08} examines up to 3553 half-lives of nuclei with various decay modes.
The situation does not change if we add a few hundred or thousand estimated half-life values to the
number sequence \cite{NiRen08, NiRen09}.

\par
\begin{figure}
\centering{
\resizebox{0.7\textwidth}{!}{
  \includegraphics{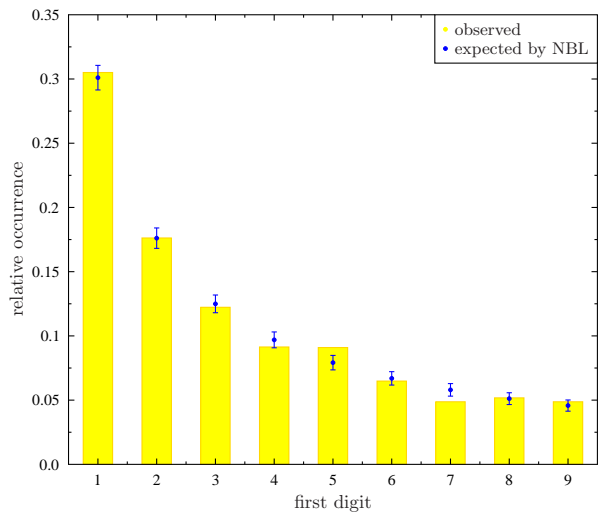}
}}
\caption{The distribution of the first significant digit of half-lives follows the Newcomb\,--\,Benford law. 2298 half-life
values from NUBASE2003 \cite{NUBASE} had been evaluated.}
\label{fig:benford}
\end{figure}

 As an illustration, we show the result of our own analysis in Table~\ref{tab:1} and in Fig.~\ref{fig:benford}.
The input data sequence had been taken from NUBASE2003 \cite{NUBASE}. The nuclei for which we only know the 
upper or lower limit of their half-life values had been rejected, resulting in an input sequence of 2298 entries. The
occurrence of the first digits follow the NBL well within two standard deviations.

\begin{table}
\caption{The occurrence of the first significant digit of half-lives is in good agreement with the Newcomb\,--\,Benford law.}
\label{tab:1}
\center{
\begin{tabular}{ccc}
\hline\noalign{\smallskip}
Digit & Occurrence & Expected by the NBL\\
\noalign{\smallskip}\hline\noalign{\smallskip}
1 & 701 & 692 $\pm$ 22 \\
2 & 405 & 405 $\pm$ 18 \\
3 & 281 & 287 $\pm$ 16 \\
4 & 210 & 223 $\pm$ 14 \\
5 & 209 & 182 $\pm$ 13 \\
6 & 149 & 154 $\pm$ 12 \\
7 & 112 & 133 $\pm$ 11 \\
8 & 119 & 118 $\pm$ 11 \\
9 & 112 & 105 $\pm$ 10 \\
\noalign{\smallskip}\hline
\end{tabular}
}
\end{table}

\par The calculation of the error of the NBL is based on the binomial distribution \cite{Buck93}. Let $N$ be
the size of the data set (in our case $N = 2298$) and $P_d$ the probability of a value having the digit $d$ as its
first significant digit. One is expected to find $N_d = NP_d$ values out of $N$ having $d$ as its first significant
digit, with an error of $\Delta[N_d]$, where
\begin{equation}
	\label{eq:error}
	\Delta[N_d] = \sqrt{ N P_d [1 - P_d] }
\end{equation}
according to the standard deviation of the binomial distribution.

\subsection{Ones scaling test} \label{sec:2.2}

\par
\begin{figure}
\centering{
\resizebox{0.7\textwidth}{!}{
  \includegraphics{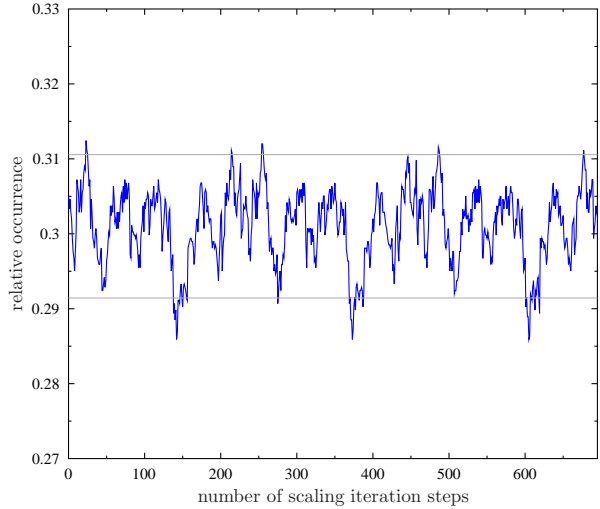}
}}
\caption{The result of the ones scaling test executed on our data set. The relative number of ones as the first
significant digit scatters around its expected value $0.301$ within the error $0.0096$. The scaling constant is
$1.01$. The error is given by Eq.\,\ref{eq:error}, and indicated by the horizontal lines.}
\label{fig:ost}
\end{figure}

 The `ones scaling test' (OST) is a simple method to test quantitatively whether a data set satisfies the Newcomb\,--\,Benford
law \cite{DSPBook}. The data set obeys the law if the relative occurrence of the numbers beginning with the digit 1 is
around $\mathrm{P}_1 = 30.1$\,\%, even after an arbitrary number of multiplications by a given constant (scaling
invariance). The used constant is the scaling constant, which is $1.01$ in our case. Figure\,\ref{fig:ost} shows the
result of the OST for the 2298 half-lives we examined. It is clear that the NBL is followed quite precisely.

\section{The mathematical background of the Newcomb\,--\,Benford law} \label{sec3}

\par In this section we summarize the conditions when the NBL is satisfied, based on the work of \emph{Smith}
\cite{DSPBook}. The following reasoning can be used whenever one knows the probability density function (p.\,d.\,f.)\
of the set of numbers for which the compliance of the NBL is investigated.

\par
\begin{figure}
\centering{
\resizebox{0.7\textwidth}{!}{
  \includegraphics{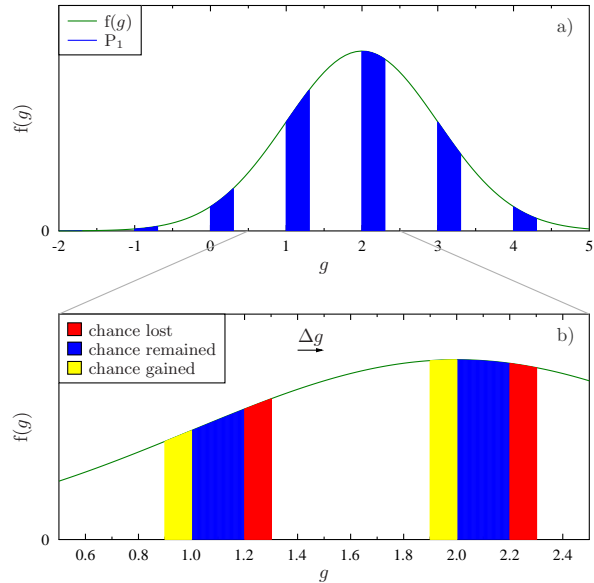}
}}
\caption{Illustration of the sampling of the ones scaling test: \emph{a)} an example distribution
(log-normal with a mean of 2 and standard deviation 1 on the lg axis), \emph{b)} the result of the multiplication
of the number sequence by $10^{\Delta g}$.}
\label{fig:sampling}
\end{figure}

 A data set obeys the NBL if and only if
its ones scaling test gives $\lg2 \approx 0.301$ after
any number of multiplications by the scaling constant. Let us switch from linear scale to logarithmic one:
$g(x) = \lg x$, and let the set of numbers have a p.\,d.\,f.\ $\mathrm{f}(g)$. The condition that a number
has $1$ as its first significant digit can be
described with a sampling function $\mathrm{s}(g)$, which is a periodic function (a square wave) on the logarithmic
scale. Then the $\mathrm{P}_1$ probability of having $1$ as the first significant digit is (Fig.\,\ref{fig:sampling})
\begin{equation}
	\label{eq:P1}
  	\mathrm{P}_1 = \int\limits_{-\infty}^\infty \mathrm{s}(g)\mathrm{f}(g)\mathrm{d}g.
\end{equation}

\par Multiplication by $c^m$ (where $c > 1$ is the scaling constant) works as a shift by $\gamma = m \lg c$ on
the logarithmic scale. The area denoted by $\mathrm{P}_1$ in Eq.\,\ref{eq:P1} remains unchanged if
we gain and lose the same area when we perform a scaling (Fig.\,\ref{fig:sampling}) \cite{Fewster09}. Due to the
shape of the sampling function, a scaling by $10^i\ (i \in \mathbb{Z})$ results in the same
$\mathrm{P}_1$, making the results of the OST necessarily periodic when $c \approx 1$ (as in Fig.\,\ref{fig:ost}).
According to Newcomb \cite{Newcomb},
\[ \forall c \in \mathbb{R}, c > 0: \exists i \in \mathbb{Z}, r \in [0,1): c = 10^{i+r}. \]
Only $r$ plays role in the value of the first significant digit, since $i$ only represents the shifting of the
decimal point. Newcomb stated that the NBL is naturally followed, because the distribution of $r$ is uniform.
Though the uniformity of $r$ is not at all trivial, by using the train of thought above we can see, that his
statement can be written in a more precise form: the NBL is followed if the distribution of the numbers are
log-uniform ranging between integer powers of $10$.

\par To get all functions approximately obeying the law in a mathematically rigorous way, the result
of the ones scaling test after $m$ scaling iterations has to be expressed as a convolution:
\[ \mathrm{P}_1(\gamma) = \int\limits_{-\infty}^\infty \mathrm{s}(g)\mathrm{f}(g-\gamma)\mathrm{d}g =
\mathrm{s}(g) \ast \mathrm{f}(-g). \]
The convolution transforms into a simple multiplication $\mathrm{F}(f) \cdot \mathrm{S}(f)$ in the frequency domain,
where $\mathrm{F}$ and $\mathrm{S}$ are the Fourier transforms of $\mathrm{f}$ and $\mathrm{s}$, respectively.
Since the support of $\mathrm{S}(f)$ is the nonzero integer numbers, and the ones scaling test can give a constant
$0.301$ function only if the $\mathrm{F}(f) \cdot \mathrm{S}(f)$ product is $0.301$ at $f = 0$ and vanishes
everywhere else, the following theorem stands:
A set of numbers with the probability distribution $\mathrm{f}(g)$ obeys the NBL if and only if the
Fourier transform $\mathrm{F}(f)$ of the probability distribution function of the numbers vanishes at all nonzero
integer frequencies (Benford's law compliance theorem, \cite{DSPBook}). For an experimentally good satisfaction of
the law, it is enough for the product to become small at integer frequencies. In natural sciences this criterion is
most likely to be fulfilled by such $\mathrm{f}(g)$ distributions the $\mathrm{F}(f)$ Fourier transform of which
becomes very small ($\mathrm{F}(f) \approx 0$) before $f = 1$, and remains close to zero thereafter.


\section{Conclusions} \label{sec4}

\par The specific distribution of the first digits comes from the method one
uses when decides whether a number has $d$ as its first digit. This method is mathematically defined by the sampling
function. It follows, that for any given distribution of numbers, the compliance of
the NBL (within a given error) can be predicted, but not vice versa: from the compliance of the law the distribution cannot be
constructed. 

\par Consequently (and contrary to the findings of \cite{NiRen08, NiRen09}), the feeling that there is a natural phenomenon behind the NBL is illusoric. While scale
invariant systems play an important role in today's research, in the case of the NBL the source of the scale
invariance is the periodicity of the sampling function (given the probability distribution fulfills the mentioned
criterion).

\par If one knows the distribution of a number set, the NBL carries no additional information. Thus, it is only the
distribution that should be explained or predicted by the theory of a given phenomena obeying the law, and not
the satisfaction of the law itself. Contrary to the suggestion of e.\,g.\ \cite{Pain08, NiRen08, NiRen09}, knowing the p.\,d.\,f.,\ the
use of the NBL to test a number sequence derived from a theory modeling a phenomenon that obeys the law gives
no additional information on the physics of the system (just as knowing the integral of a function on an interval does not help much in finding a function with a given shape).

\par
\begin{figure}
\centering{
\resizebox{0.7\textwidth}{!}{
  \includegraphics{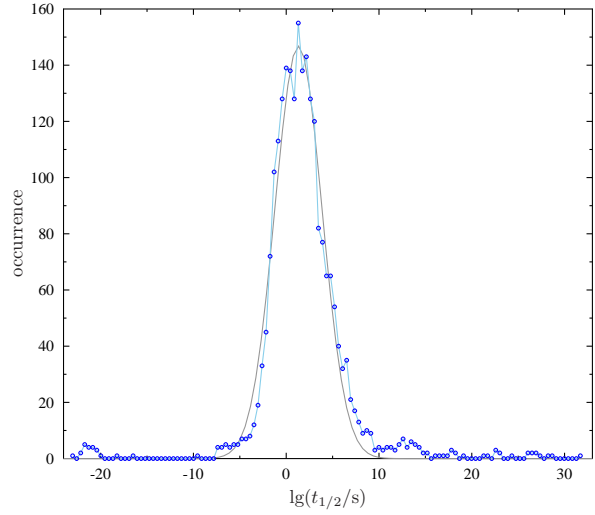}
}}
\caption{Distribution of the half-lives of the examined 2298 decays. The known half-life values cover about 54 orders
of magnitude. The binning of the histogram is natural logarithm based.}
\label{fig:distrib}
\end{figure}

 In the case of nuclear decay, the distribution of the half-lives is known (see Fig.\,\ref{fig:distrib}).  The
examined experimental values of the half-lives span $54$ decimal orders of magnitude. It can be seen that this
distribution is very close to a log-normal distribution, which may suggest a multiplicative process in the
background. The parameters of the fitted normal distributions can be read in Table \ref{tab:2}.

\begin{figure}
\centering{
\resizebox{0.7\textwidth}{!}{
  \includegraphics{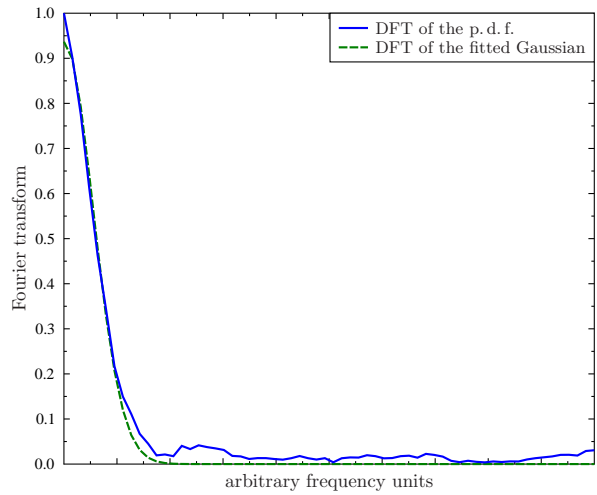}
}}
\caption{Fourier transform of the half-life distribution and the fitted log-normal function.}
\label{fig:match}
\end{figure}

By looking at the
distribution it is not surprising that half-lives follow the NBL. The Fourier transform of
a Gaussian (see Fig.\,\ref{fig:match}) with standard deviation $\sigma$ is yet another Gaussian with a mean of $0$
and a standard deviation of
$1/(2\pi\sigma)$. The standard deviation of the fitted Gaussian is $3.0139/\sqrt{2\ln 2} \approx 2.56$ on the
decimal logarithmic scale, making the standard deviation of its Fourier transform $\approx 0.062$. Thus, the
transform becomes very small at all integer frequencies, resulting only minor fluctuations in the value of
$P_1(\gamma)$.

\par If a theory of nuclear decay describes this distribution well, it will automatically reproduce the Newcomb\,--\,Benford law.
Thus, the physics is behind the form of the distribution, and the models have nothing to do with the Newcomb\,--\,Benford law itself.

\begin{table*}
\caption{The parameters of the Gaussians fitted to the distribution of the half-lives of 2298 decays. The parameters
differ slightly depending on the resolution of the binning. Decimal, natural and binary logarithm based binning has
been used.}
\label{tab:2}
\center{
\begin{tabular}{ccccc}
\hline\noalign{\smallskip}
binning & mean & $\mathrm{base}^\mathrm{mean}$ & HWHM & $\mathrm{base}^\mathrm{HWHM}$\\
\noalign{\smallskip}\hline\noalign{\smallskip}
lg & $1.0503 \pm 0.0051$ & $11.23$ & $3.0139 \pm 0.0060$ & $1032.52$ \\
ln & $3.030 \pm 0.018$ & $20.70$ & $6.889 \pm 0.021$ & $981.42$\\
lb & $4.609 \pm 0.030$ & $24.40$ & $9.909 \pm 0.036$ & $961.40$\\
\noalign{\smallskip}\hline
\end{tabular}
}
\end{table*}


\section*{Acknowledgments}
\par This work was supported by OTKA (K68801).


\end{document}